\documentclass[12pt]{article}

\usepackage{amsmath} 
\usepackage{amsthm,amsmath,mathrsfs,amsfonts,amssymb}
\usepackage{graphicx,psfrag,epsf,fixmath,caption,subcaption}
\usepackage{enumerate}
\usepackage{natbib} \RequirePackage[colorlinks,citecolor=blue,urlcolor=blue,linkcolor=blue]{hyperref}
\usepackage{url} 

\usepackage{enumitem} 

\newcommand{\blind}{1}
\theoremstyle{definition}

\begin{document}

\def\spacingset#1{\renewcommand{\baselinestretch}{#1}\small\normalsize} \spacingset{1}

\if1\blind
{
\title{\bf {\normalsize PEER EFFECTS AND ENDOGENOUS SOCIAL INTERACTIONS}}
\author{ {\small KOEN JOCHMANS}\thanks{Address: University of Cambridge, Faculty of Economics, Austin Robinson Building, Sidgwick Avenue, Cambridge CB3~9DD, United Kingdom. E-mail: \texttt{kj345@cam.ac.uk}. Financial support from the European Research Council through grant n\textsuperscript{o} 715787 (MiMo) is gratefully acknowledged. This version is dated \today.}\\ {\small UNIVERSITY OF CAMBRIDGE}
}
\date{} 
  \maketitle
} \fi

\if0\blind
{
  \bigskip
  \bigskip
  \bigskip
  \begin{center}
    {\bf {\normalsize }}
\end{center}
  \medskip
} \fi

\medskip
\vspace{-.75cm}

\begin{abstract}
\noindent
We introduce an approach to deal with self-selection of peers in the linear-in-means model. Contrary to the existing proposals we do not require to specify a model for how the selection of peers comes about. Rather, we exploit two restrictions that are inherent to many such specifications to construct intuitive instrumental variables. These restrictions are that link decisions that involve a given individual are not all independent of one another, but that they are independent of the link behavior between other pairs of individuals. We construct instruments from the subnetwork obtained on leaving-out all one's own link decisions in a manner that is reminiscent of the approach \cite{BramoulleDjebbariFortin2009} when the assignment of peers is assumed exogenous. A two-stage least-squares estimator of the linear-in-means model is then readily obtained. 

\bigskip
\noindent
{\bf Keywords:}  
instrumental variable,
network,
self-selection

\medskip
\noindent
{\bf JEL classification:}
C31,
C36

\end{abstract}


\spacingset{1.45}

\renewcommand{\theequation}{\arabic{section}.\arabic{equation}} 
\setcounter{equation}{0}


\newpage

\section*{Introduction}
The importance of acknowledging the existence of social interactions between agents in the estimation of causal relationships is now widely acknowledged. In a program-evaluation problem, for example, non-treated individuals can nonetheless benefit from the program through spillovers from treated units with whom they interact. Examples of this are detailed in \cite{MiguelKremer2004}, \cite{Sobel2006}, and \cite{AngelucciDeGiorgi2009}. A key concern when estimating models that feature peer effects is that agents may self-select their peers, and do so based on (unobserved) factors that equally feature in the equation of interest, thus creating an endogeneity problem. Randomized assignment to peer groups has proven useful in circumventing this threat to identification (\citealt{Sacerdote2001} contains any early application of this strategy) but this is, of course, not possible in many situations. The literature has worked on approaches to deal with the self-selection problem in the linear-in-means model of social interactions as introduced in \cite{Manski1993} and analyzed by  \cite{BramoulleDjebbariFortin2009} for the case where individuals interact through a general, but fixed, network. The current paper is an addition to this growing body of work.

We consider a setting where data on a large number of networks is available. This is the conventional viewpoint.\footnote{\citet[p.~537]{Manski1993} provides a discussion on the incompatibility of the model with data obtained from randomly sampling individuals. Work on inference under snowball sampling, albeit in a different setting, is in \cite{Leung2019}.} \cite{GoldsmithPinkhamImbens2013}, \cite{QuLee2015}, and \cite{HsiehLee2016} proposed control-function approaches to deal with endogeneity of the network. While the details of each of the proposals are different, the idea is to complement the linear-in-means model with a full specification of the network-formation process. The chief limitation of such an approach lies in the fact that it is not robust to misspecification of the link-formation process. In the alternative setting where a single large network is observed, \cite{Auerbach2019} and, concurrently, \cite{JohnssonMoon2019} developed more flexible, semiparametric, control-function approaches. These require a less stringent specification of the network formation but do require the network in question to grow in size and to be dense for the estimator constructed from it to be consistent. Therefore, they cannot be modified to fit our sampling paradigm of many networks. At the same time, they also maintain the requirement that the unobserved heterogeneity driving the network endogeneity is univariate.

Here, we take an instrumental-variable route to deal with the endogeneity of the peer group. The advantage of such a strategy is that it does not require a tight specification of the model driving network formation, allowing for the source of the endogeneity problem to be multidimensional, for example. In related work, \cite{KelejianPiras2014}, and later also \cite{LeeLiuPatacchiniZenou2020}, equally considered an instrumental-variable approach, regressing link outcomes on exogenous variables that are presumed to drive the link decisions to cleanse them from endogenous factors. While this is a general and simple technique, the link predictions so constructed will tend to be poor predictors of actual link decisions unless the latter are mainly driven by the exogenous variables in question. A discussion on this is in \cite{LeeLiuPatacchiniZenou2020}, and we equally observed this in our simulation work.

We, in stead, exploit two restrictions on network formation that are implicit in most network-formation models investigated in the literature to generate instrumental variables that are internal to the model, in the same vain as in a dynamic panel data model. These restrictions are that (i) link decisions of a given individual are dependent, but that (ii) link decisions involving any two distinct pairs of agents are (conditionally) independent. Condition (ii) limits the degree of the endogeneity problem. In turn, the implication of Condition (i) is that link decisions between any triple of individuals are informative about each other. Together, these conditions pave the way for the construction of instrumental variables. They are sufficiently general to cover both settings where networks are formed cooperatively or non-cooperatively, allow for the possibility of transfers between individuals, and accommodate homophily of unrestricted form, for example. They are satisfied in the models of \cite{Auerbach2019} and \cite{JohnssonMoon2019}, for example, as well as in more general versions thereof. Importantly, our conditions do rule out interdependent link formation behavior such as transitivity, where individuals are more likely to link if they have more connections in common. This is equally ruled-out in the control-function approach as, there, such a mechanism would result in an incomplete model, rendering the parameters of the network-formation model set- as opposed to point-identified, causing the approach to break down.

In the linear-in-means model the outcome of a given individual depends on the average outcome and the average characteristics of her peers, as well as on her own characteristics. When peer groups are exogenous only the first of these peer effects creates an endogeneity problem. The approach of  \cite{BramoulleDjebbariFortin2009}, in essence, instruments the average peer outcome by the average characteristics of the peers of peers. We, in stead, are faced with a situation in which both types of peer effect are endogenous. We  construct instrumental variables as follows. For each individual we set up the subnetwork obtained on removing all links in which this individual is involved. Under our conditions this leave-own-out network is exogenous and contains useful predictive information about the individuals own link behavior. Next, we instrument average peer characteristics by the average of these characteristics in the leave-own-out network. In the same way, we instrument average peer outcomes by the average of the characteristics of peers of peers in the leave-own-out network. Like in the exogenous case, the procedure can be iterated to involve characteristics of peers further away in the network. This is an intuitive extension of \cite{BramoulleDjebbariFortin2009}. The resulting two-stage least-squares procedure is standard to implement and generates the usual procedures to test for network endogeneity through a Durbin-Wu-Hausman test. The estimator's asymptotic distribution follows from \cite{HansenLee2019}.

Below we first set up the model, state our identifying assumptions, and motivate them by showing how they are implied in the setting of \cite{Auerbach2019} and \cite{JohnssonMoon2019}. We then give our instrumental variables as derived from the leave-own-out networks and next present the resulting two-stage least-squares estimator along with its large-sample distribution. Results from a simulation experiment show that it performs well in small samples.

\section{Setup} \setcounter{equation}{0}
Our asymptotics will involve data on many networks but, for now, it suffices to consider a single network. 


\paragraph{Model}
Consider an undirected network involving $n$ agents. Let $\boldsymbol{A}$ denote its $n\times n$ adjacency matrix. Then
$$
(\boldsymbol{A})_{i,j}
=
\left\lbrace
\begin{array}{cl}
1 & \text{ if } i \text{ and } j \text{ are connected} \\
0 & \text{ otherwise}	
\end{array}	
\right. .
$$  
When a link exists between $(i,j)$ we say that they are neighbors. As usual, we do not consider agents to be linked with themselves, so matrix $\boldsymbol{A}$ has only zeros on its main diagonal. It will be useful to have a notational shorthand for the row-normalized adjacency matrix,  $\boldsymbol{H}$, say. Its entries are
$$
(\boldsymbol{H})_{i,j}
=
\left\lbrace
\begin{array}{cl}
{(\boldsymbol{A})_{i,j}} \left/ {\sum_{j^\prime=1}^n (\boldsymbol{A})_{i,j^\prime}} \right. & \text{ if }	\sum_{j^\prime=1}^n (\boldsymbol{A})_{i,j^\prime}> 0 \\
0 & \text{ otherwise }
\end{array}	
\right. .
$$
Recall that $\boldsymbol{H}$ corresponds to the transition matrix of a random walk through our network. Moreover, $(\boldsymbol{H})_{i,j}$ is the probability that, when taking a single step, starting at agent $i$, we arrive at agent $j$. In the same way, $(\boldsymbol{H}^2)_{i,j}$ is the probability of arriving in two steps, and so on.

Let $y_i$ and ${x}_i$ denote scalar variables, observable for each agent. Our baseline model is 
$$
y_i = \alpha +  {\beta} {x}_i  +  \gamma
{\textstyle \left(\sum_{j=1}^n (\boldsymbol{H})_{i,j}{x}_j \right) }
+ \varepsilon_i,
$$
where $\varepsilon_i$ is a mean-zero unobserved variable. Taking the regressor to be a scalar is done only for notational convenience.
Here, ${\beta}$ captures the direct effect of ${x}_i$ on $y_i$ while ${\gamma}$ reflects an indirect, spillover, effect from the covariate values of the neighbors.
In matrix form we can succinctly write
$$
\boldsymbol{y} = \alpha\boldsymbol{\iota}_n + {\beta}\boldsymbol{x} + {\gamma}\boldsymbol{H}\boldsymbol{x} + \boldsymbol{\varepsilon},
$$
where $\boldsymbol{y}=(y_1,\ldots,y_n)^\prime$, $\boldsymbol{\iota}_n=(1,\ldots,1)^\prime$ is the $n$-vector of ones,  $\boldsymbol{x}=({x}_1,\ldots, {x}_n)^\prime$, and $\boldsymbol{\varepsilon}=(\varepsilon_1,\ldots, \varepsilon_n)^\prime$.
An extension of the baseline specification that accommodates endogenous peer effects, where $y_i$ also depends on $\sum_{j=1}^n (\boldsymbol{H})_{i,j} y_j$, gives rise to what we will call the full model, 
$$
\boldsymbol{y} = \alpha\boldsymbol{\iota}_n + \delta \boldsymbol{H}\boldsymbol{y}  + {\beta}\boldsymbol{x} + {\gamma}\boldsymbol{H}\boldsymbol{x} + \boldsymbol{\varepsilon},
$$
which is the workhorse linear-in-means model on general networks as studied in \cite{BramoulleDjebbariFortin2009} and \cite{DeGiorgiPellizzariRedaelli2010}. It will be useful to first present our approach in the baseline model. The extension to the full model will then be intuitive.


\paragraph{Restrictions}
Identification of the slope coefficients in our model is well-understood when the strict exogeneity condition 
$
\mathbb{E}(\varepsilon_i \vert \boldsymbol{A}, \boldsymbol{x}) = 0 \ \text{(a.s.)}
$
holds. Here we relax this restriction by allowing for dependence between the link decisions and the unobserved component in our model. We work with
\begin{equation} \label{eq:restriction}
\mathbb{E}(\varepsilon_i \vert \boldsymbol{A}_{i}, \boldsymbol{x}) = 0 \ \text{(a.s.)},
\end{equation}
where $\boldsymbol{A}_{i}$ is the $(n-1)\times (n-1)$ adjacency matrix of the subnetwork obtained from $\boldsymbol{A}$ on deleting its $i$th row and its $i$th column. This condition implies unconditional moments that can be used in a two-stage least-squares procedure. For our instruments to be relevant we will presume that 
\begin{equation} \label{eq:dependence}
\mathbb{E}((\boldsymbol{A})_{i,j} \vert 
\boldsymbol{A}_i, \boldsymbol{x})
\neq 
\mathbb{E}((\boldsymbol{A})_{i,j} \vert \, \boldsymbol{x}) \ \text{(a.s.)}.
\end{equation}
This condition states that the link decisions of a given agent are not independent of one another, conditional on the covariate, and is natural in our context. Before turning to our instrumental-variable approach we provide motivation and justification for the conditions in \eqref{eq:restriction} and \eqref{eq:dependence}.

\paragraph{Network formation}
It is useful to start with the model for link decisions specified in \cite{Auerbach2019} and \cite{JohnssonMoon2019}. They stipulate that, for each pair of agents $(i,j)$, with $i < j$, 
\begin{equation} \label{eq:stylized}
(\boldsymbol{A})_{i,j}
=
(\boldsymbol{A})_{j,i}
=
\left\lbrace
\begin{array}{cl}
1 & \text{ if } h(\eta_i,\eta_j)> u_{i,j}
\\
0 & \text{ otherwise }	
\end{array}	
\right.  ,
\end{equation}
where $\eta_1,\ldots,\eta_n$ are independent scalar random variables, the $u_{i,j}$ are random shocks, and $h$ is a conformable function.\footnote{\cite{Auerbach2019} and \cite{JohnssonMoon2019} also impose certain shape restrictions on the function $h$ and an i.i.d.~assumption on the $u_{i,j}$ to achieve identification in their setting but these are not important for our developments and, thus, not imposed here.} Link decisions are allowed to be endogenous because $\eta_i$ and $\varepsilon_i$ are allowed to be dependent. The shocks $u_{i,j}$ are independent of $(\eta_i,\eta_j)$ and of $(\varepsilon_i,\varepsilon_j)$ The implication is that
$$
\mathbb{E}(\varepsilon_i \vert \boldsymbol{A}, \boldsymbol{x})
=
\mathbb{E}(\varepsilon_i \vert \eta_i, \boldsymbol{x})
=
\mathbb{E}(\varepsilon_i \vert \eta_i).
$$
Because $(\boldsymbol{A})_{i,j}$ depends on $(\eta_i,\eta_j)$ all link decisions involving agent $i$ correlate with $\varepsilon_i$.
However, for all $i^\prime\neq i$ and $j\neq i$,  the link decision $(\boldsymbol{A})_{i^\prime,j}$ is independent of $\varepsilon_i$. Consequently, 
$$
\mathbb{E}(\varepsilon_i \vert \boldsymbol{A}_i, \boldsymbol{x})
=
\mathbb{E}(
\mathbb{E}(\varepsilon_i \vert \boldsymbol{A}, \boldsymbol{x}) \vert  \boldsymbol{A}_i, \boldsymbol{x})
=
\mathbb{E}(\mathbb{E}(\varepsilon_i \vert \eta_i) \vert \eta_1,\ldots,\eta_{i-1},\eta_{i+1},\ldots,\eta_n) = \mathbb{E}(\varepsilon_i) = 0,
$$
meaning that our moment restriction in \eqref{eq:restriction} holds in the model of \cite{Auerbach2019} and \cite{JohnssonMoon2019}. Furthermore, for all $i^\prime\neq i$, $(\boldsymbol{A})_{i,j}$ and $(\boldsymbol{A})_{i^\prime,j}$ are dependent as they are both functions of $\eta_j$. Hence, there is predictive information about the former in the latter, and \eqref{eq:dependence} is satisfied.

The stylized model in \eqref{eq:stylized} can be generalized in a number of ways without jeopardising the validity of \eqref{eq:restriction} and \eqref{eq:dependence}. The features that are embedded in it that are important for our purposes are that (i) $\eta_i$ only affects link decisions involving agent $i$; (ii) the  $\eta_1,\ldots, \eta_n$ are independent conditionally on the ${x}_1,\ldots, {x}_n$; and (iii) the decision to link depends on characteristics of both agents involved. These restrictions allow $\eta_i$ to be replaced by a vector of agent-specific unobserved heterogeneity and permit link decisions to depend on a set of additional (observable or unobservable) variables. The function $h$ could also be allowed to be pair-specific. These generalizations allow for excess heterogeneity and homophily of unrestricted form. 

The specification in \eqref{eq:stylized} is suitable when link formation is cooperative. A modified version of it, suitable for non-cooperative situations, would be
\begin{equation} \label{eq:stylized2}
(\boldsymbol{A})_{i,j}
=
(\boldsymbol{A})_{j,i}
=
\left\lbrace
\begin{array}{cl}
1 & \text{ if } h(\eta_i)> u_{i,j} \text{ and }  h(\eta_j)> u_{j,i}
\\
0 & \text{ otherwise }	
\end{array}	
\right.  ,
\end{equation}
where, now, $u_{i,j}$ and $u_{j,i}$ are pair-specific shocks. Clearly, such a specification also satisfies all requirements for \eqref{eq:restriction} and \eqref{eq:dependence} to hold. Again, \eqref{eq:stylized2} can be extended in a variety of ways without compromising this.

This discussion shows that \eqref{eq:restriction} is implied by many commonly-used specifications for network formation. The most important limitation of the requirements in (i)--(iii) is that they rule out situations where link decisions are interdependent. Transitivity, for example, where a pair of agents are more likely to be linked when they have more neighbors in common, calls for a simultaneous-equation model. Such a design would  violate (i), as $(\boldsymbol{A})_{i,j}$ will generally depend on all $\eta_1,\ldots,\eta_n$ in such a case. Without access to panel data, as in \cite{GoldsmithPinkhamImbens2013}, for example, dealing with such a design appears complicated.

\section{Approach} \setcounter{equation}{0}

Start with the baseline model. Here, the self-selection of ones' peers causes the spillover effect $\sum_{j=1}^n (\boldsymbol{H})_{i,j} x_j$ to be endogenous because the weights $(\boldsymbol{H})_{i,1},\ldots,(\boldsymbol{H})_{i,n}$ correlate with the unobserved component $\varepsilon_i$. Because $\boldsymbol{H}$ is a row-normalized adjacency matrix, $(\boldsymbol{H})_{i,j}$ depends on all of $(\boldsymbol{A})_{i,1},\ldots,(\boldsymbol{A})_{i,n}$. It does not depend on $(\boldsymbol{A})_{i^\prime,j^\prime}$ for any of $i^\prime\neq i$ and $j^\prime\neq i$, however. By \eqref{eq:restriction}, the link decisions that do not involve agent $i$ are exogenous. Furthermore, by \eqref{eq:dependence}, these $(\boldsymbol{A})_{i^\prime,j^\prime}$ are not independent of $(\boldsymbol{H})_{i,j}$. This suggests the construction of instrumental variables by looking at linear combinations of $x_1,\ldots,x_n$, with weights coming from the leave-one-out network $\boldsymbol{A}_i$. A fruitful way of doing so is discussed next.

For each $i^\prime$ we define the $n\times n$ matrix
$$
(\boldsymbol{H}_{i^\prime})_{i,j}
=
\left\lbrace
\begin{array}{cl}
{(\boldsymbol{A})_{i,j}} \left/ {\sum_{j^\prime \neq i^\prime}  (\boldsymbol{A})_{i,j^\prime}}	\right.
& \text{ if } i\neq i^\prime \text{ and } j\neq i^\prime \text{ and } \sum_{j^\prime \neq i^\prime} (\boldsymbol{A})_{i,j^\prime} > 0
\\
0 & \text{ otherwise}
\end{array}	
\right. .
$$
This is the row-normalized version of the adjacency matrix $\boldsymbol{A}_{i^\prime}$ introduced previously, only complemented with one additional zero row and one additional zero column. This augmentation is done for notational considerations, as it maintains the dimension of these matrices to $n\times n$. We stress that $\boldsymbol{H}_i$ is not obtained from setting to zero the corresponding row and column of $\boldsymbol{H}$. We can interpret $\boldsymbol{H}_{i}$ as the transition matrix on the network obtained on ruling-out links that involve agent $i$.
From \eqref{eq:restriction}, the entries of this matrix are uncorrelated with $\varepsilon_{i}$. Furthermore, from \eqref{eq:dependence}, $(\boldsymbol{H}_i)_{i^\prime,j^\prime}$ and $(\boldsymbol{H})_{i,j}$ are dependent for all $(i,j)$ and $(i^\prime,j^\prime)$, conditional on the regressors.

Recall that $(\boldsymbol{H})_{i,j}$ is the probability of arriving at agent $j$, from agent $i$, in a single step in the network defined by the original adjacency matrix $\boldsymbol{A}$. The entries of the $n\times n$ matrix
$$
(\boldsymbol{Q}_1)_{i,j}
=
{\textstyle
(n-1)^{-1} \sum_{i^\prime \neq i} (\boldsymbol{H}_i)_{i^\prime,j}},
$$
in contrast, give the probability of arriving at agent $j$ in the network defined by $\boldsymbol{A}_i$, no matter the starting point, in a single step. 
The average 
$
\textstyle
\sum_{j=1}^n (\boldsymbol{Q}_1)_{i,j} x_j
$
is exogenous and will correlate with the spillover term, $\sum_{j=1}^n (\boldsymbol{H})_{i,j} x_j$. A simple intuition can be given by considering the example of network centrality: if agent $j$ is involved in many links, it is likely that she will also be linked with agent $i$. We may then predict the link decision between agents $i$ and $j$ by looking at the linking behavior of agent $j$ with all the other agents in the network. These choices are exogenous. 
A two-stage least-squares approach has just suggested itself for the baseline model: we instrument the endogenous spillover effect $\boldsymbol{H}\boldsymbol{x}$ by $\boldsymbol{Q}_1\boldsymbol{x}$. The rank condition for identification here requires that $\boldsymbol{x}$ is not proportional to the unit vector $\boldsymbol{\iota}_n$ and that $\boldsymbol{Q}_1\boldsymbol{x}$ covaries with $\boldsymbol{H}\boldsymbol{x}$, after $\boldsymbol{\iota}_n$ and $\boldsymbol{x}$ have been projected-out from it. This last condition is, of course, simply the usual relevance condition and requires that $\boldsymbol{Q}_1\boldsymbol{x}$ has some predictive power for $\boldsymbol{H}\boldsymbol{x}$ after controlling for $\boldsymbol{x}$. It appears difficult to come up with situations where this will fail---provided, of course, that \eqref{eq:dependence} holds---except for trivial cases. The empty network and the complete network are two such examples; clearly, in these examples identification would also fail under network exogeneity; $\boldsymbol{H}\boldsymbol{x}$ is a linear function of $\boldsymbol{\iota}_n$ and $\boldsymbol{x}$, leading to a standard multicolinearity problem. Other instruments than $\boldsymbol{Q}_1\boldsymbol{x}$ can equally be constructed under our exclusion restriction; some examples follow below. The current choice, however, has a natural extension to the full model, to which we turn next.

%

In the full linear-in-means model,
$$
\boldsymbol{y} = \alpha\boldsymbol{\iota}_n + \delta \boldsymbol{H}\boldsymbol{y}  + {\beta}\boldsymbol{x} + {\gamma}\boldsymbol{H}\boldsymbol{x} + \boldsymbol{\varepsilon},
$$
the presence of $\boldsymbol{H}\boldsymbol{y}$ as a regressor would induce an endogeneity problem even if $\boldsymbol{H}$ were exogenous. If $-1<\delta <1$, and if all the agents in the network are linked to at least one other agent, 
\begin{equation} \label{eq:reduced}
\boldsymbol{H}
\boldsymbol{y}
=
\mu \, \boldsymbol{\iota}_n  
+
\beta \boldsymbol{H}\boldsymbol{x}
+
\lambda
\sum_{s=0}^\infty \delta^s \boldsymbol{H}^{s+2}\boldsymbol{x}
+
\sum_{s=0}^\infty \delta^s \boldsymbol{H}^{s+1}\boldsymbol{\varepsilon},
\end{equation}
where we write $\mu = \alpha/(1-\delta)$ and $\lambda = \delta\beta + \gamma$. The argument of \cite{BramoulleDjebbariFortin2009} and \cite{DeGiorgiPellizzariRedaelli2010} is that $\boldsymbol{H}^2\boldsymbol{x}$, $\boldsymbol{H}^3\boldsymbol{x}$, and so on can be used as instrumental variables for $\boldsymbol{H}\boldsymbol{y}$ when the network is exogenous, provided that $\lambda \neq 0$. The validity of these variables as instruments breaks down when the network is endogenous.

On inspecting the expansion in \eqref{eq:reduced} a natural extension to our approach in the baseline model presents itself. 
Because $\boldsymbol{H}_i$ is a matrix of transition probabilities, it can be iterated on, in the same way as $\boldsymbol{H}$, to yield probabilities of arriving at each agent when taking multiple steps through the network. 
In full analogy to $\boldsymbol{Q}_1$, the entries of the $n\times n$ matrix
$$
(\boldsymbol{Q}_2)_{i,j}
=
{\textstyle
(n-1)^{-1} \sum_{i^\prime \neq i} \sum_{j^\prime=1}^n (\boldsymbol{H}_i)_{i^\prime,j^\prime} \, (\boldsymbol{H}_i)_{j^\prime,j}},
$$
give the probability of arriving at agent $j$ in the network induced by $\boldsymbol{A}_i$, no matter the starting point, in two steps. Under our moment condition in \eqref{eq:restriction} these weights are, again, exogenous. This, then, allows to instrument the endogenous right-hand side variables, $\boldsymbol{H}\boldsymbol{x}$ and $\boldsymbol{H}\boldsymbol{y}$, by the exogenous variables $\boldsymbol{Q}_1\boldsymbol{x}$ and $\boldsymbol{Q}_2\boldsymbol{x}$. In light of the above the interpretation of this is immediate. Like in the exogenous case, we require $\boldsymbol{Q}_1$ and $\boldsymbol{Q}_2$ to be sufficiently different. Contrary to \cite{BramoulleDjebbariFortin2009}, it is more difficult here to give simple primitive conditions for instrument relevance, however. In their case \eqref{eq:reduced} implies a linear reduced form from which such conditions can be derived. This does not apply here.

Because the parameters in our model are overidentified we can consider additional instruments. One natural way to do so is by taking additional steps through the network. Letting
$$
(\boldsymbol{Q}_s)_{i,j}
=
{\textstyle
(n-1)^{-1} \sum_{i^\prime \neq i} \sum_{j_1=1}^n \cdots \sum_{j_{s-1}=1}^n (\boldsymbol{H}_i)_{i^\prime,j_1} (\boldsymbol{H}_i)_{j_1,j_2} \, \cdots \, (\boldsymbol{H}_i)_{j_{s-1},j}},
$$
for any integer $s$ it is apparent that $\boldsymbol{Q}_s\boldsymbol{x}$ is a valid instrumental variable. Instruments so constructed play a role analogous to $\boldsymbol{H}^s\boldsymbol{x}$ in the approach of \cite{BramoulleDjebbariFortin2009}.  Of course, like there, as $s$ increases the transition matrix $\boldsymbol{H}_{\hspace{-.05cm}i}^s$ will tend to its steady-state distribution, so that higher iterations will provide increasingly less (additional) information. We also note that other instruments are equally possible. For example, noting that
$$
\textstyle
\sum_{j=1}^n (\boldsymbol{Q}_s)_{i,j}\boldsymbol{x}_j =  \boldsymbol{\iota}_n^\prime\boldsymbol{H}_{\hspace{-.05cm}i}^s\boldsymbol{x}/(n-1)
$$
is an average of the vector $\boldsymbol{H}_{\hspace{-.05cm}i}^s \boldsymbol{x}$, it would be natural to consider second moments like $\boldsymbol{x}^\prime\boldsymbol{H}_{\hspace{-.05cm}i}^\prime \boldsymbol{H}_{\hspace{-.05cm}i} \boldsymbol{x}$ and $\boldsymbol{x}^\prime\boldsymbol{H}_{\hspace{-.05cm}i}^\prime \boldsymbol{H}_{\hspace{-.05cm}i}^2 \boldsymbol{x}$, and so on.

The condition that $\lambda\neq 0$ is crucial to the approach when the network is exogenous. It requires that $\boldsymbol{x}$ affects $\boldsymbol{y}$, either directly or through $\boldsymbol{H}\boldsymbol{x}$, and also that endogenous and exogenous peer effects do not exactly cancel each other out in the reduced form. Moreover, when $\lambda = 0$ we would have 
$$
\mathbb{E}(\boldsymbol{H}
\boldsymbol{y} \vert \boldsymbol{H},\boldsymbol{x}) = \mu \, \boldsymbol{\iota}_n 
+
\beta\boldsymbol{H}\boldsymbol{x}
$$ 
when link formation is exogenous; the $\boldsymbol{H}^{s}\boldsymbol{x}$, for all $s>1$, no longer contain predictive information about $\boldsymbol{H}\boldsymbol{y}$, conditional on $\boldsymbol{H}\boldsymbol{x}$. The situation is different when link formation is endogenous. Indeed, here, there will generally still be information on $\boldsymbol{H}\boldsymbol{y}$ in  $\boldsymbol{Q}_s\boldsymbol{x}$ coming from the fact that
$
\mathbb{E}(\boldsymbol{x}^\prime \boldsymbol{Q}_s^\prime \boldsymbol{H}^{p+1} \boldsymbol{\varepsilon}) \neq 0,
$
for any pair of integers $p,s$, in this case. This is so because, while the entries of $\boldsymbol{H}_{\hspace{-.05cm}i}\boldsymbol{x}$ do not correlate with $\varepsilon_i$, they do correlate with all $\varepsilon_{j}$ for $j\neq i$ when link decisions are endogenous. The implication is that endogeneity of the network can yield identification in settings where no exogenous variables are present in the linear-in-means model.

\section{Inference} \setcounter{equation}{0}
We now consider a collection of $G$ independent networks, of size $n_1,\ldots,n_G$, respectively, and rechristen $n= \sum_{g=1}^G n_g$, the total number of observations. Each of these networks comes with its associated adjacency matrix, $\boldsymbol{A}_g$---and, thus, its row-normalized version $\boldsymbol{H}_g$---as well as with the variables $\boldsymbol{y}_g=(y_{g,1},\ldots, y_{g,n_g})^\prime$ and $\boldsymbol{x}_g=(x_{g,1},\ldots, x_{g,n_g})^\prime$, which follow the linear-in-means model. We may write
$$
\boldsymbol{y}_g = \boldsymbol{X}_g \boldsymbol{\vartheta}+\boldsymbol{\varepsilon}_g,
$$
where we combine all regressors in $\boldsymbol{X}_g = (\boldsymbol{\iota}_{n_g}, \boldsymbol{H}_g\boldsymbol{y}_g, \boldsymbol{x}_g, \boldsymbol{H}_g\boldsymbol{x}_g)$ and collect all parameters to estimate in $\boldsymbol{\vartheta}=(\alpha,\delta,\beta,\gamma)^\prime$. 
The estimator of $\boldsymbol{\vartheta}$ that we consider here is the two-stage least-squares estimator using the instruments introduced above. 

On denoting the instrument matrix as $\boldsymbol{Z}_g$ the estimator can be written in its usual form
$$
\textstyle
\boldsymbol{{\vartheta}}_n
=
(\sum_g \boldsymbol{X}_g^\prime\boldsymbol{Z}_g (\sum_g \boldsymbol{Z}_g^\prime\boldsymbol{Z}_g)^{-1} \sum_g \boldsymbol{Z}_g^\prime\boldsymbol{X}_g)^{-1}
(\sum_g \boldsymbol{X}_g^\prime\boldsymbol{Z}_g (\sum_g \boldsymbol{Z}_g^\prime\boldsymbol{Z}_g)^{-1} \sum_g \boldsymbol{Z}_g^\prime\boldsymbol{y}_g).
$$
Its large-sample properties can be deduced from the work of \cite{HansenLee2019} on clustered data sets. Their results are particularly well-suited for the problem at hand. In particular, they allow for arbitrary dependence within each network, for networks to grow in size, and do not require data to be identically distributed. 
The following four conditions need to be imposed:

\medskip\noindent
(i) For some $2\leq r < \infty$, 
$$
\frac{\left(\sum_{g} n_g^r\right)^{2/r}}{n} \leq c < \infty,
\qquad
\max_g \frac{n_g^2}{n} \overset{n\uparrow \infty}{\longrightarrow}  0,
$$
where $c$ is an arbitrary constant.

\medskip\noindent
(ii) For some $s$ with $r<s$, 
$\sup_{g,i} \mathbb{E}(\lvert y_{g,i} \rvert^{2s})<\infty$ and $
\sup_{g,i} \mathbb{E}(\lvert x_{g,i} \rvert^{2s})<\infty.
$

\medskip\noindent
(iii) The matrices
$$
\frac{\sum_{g} \mathbb{E}(\boldsymbol{Z}_g^\prime\boldsymbol{\varepsilon}_g\boldsymbol{\varepsilon}_g^\prime \boldsymbol{Z}_g)}{n},
\qquad
\frac{\sum_{g} \mathbb{E}(\boldsymbol{Z}_g^\prime \boldsymbol{Z}_g)}{n}.
$$
have minimum eigenvalue bounded away from zero.

\medskip\noindent
(iv) The matrix
$$
\frac{\sum_{g} \mathbb{E}(\boldsymbol{Z}_g^\prime\boldsymbol{X}_g)}{n}
$$
has maximal column rank

\medskip\noindent
These four conditions are intuitive. A discussion on (i)--(ii) is in \cite{HansenLee2019}. Conditions (iii)-(iv) are nothing else than the usual rank conditions that are needed in any instrumental-variable problem. 

By \citet[Theorems 8 and 9]{HansenLee2019}, the estimator $\boldsymbol{\vartheta}_n$ is consistent as $n\rightarrow \infty$ and has a normal limit distribution. The robust estimator of its asymptotic variance equals
$$
\textstyle
\boldsymbol{V}_n = 
(\boldsymbol{S}_n^\prime \boldsymbol{W}_n \boldsymbol{S}_n)^{-1}
(\boldsymbol{S}_n^\prime \boldsymbol{W}_n \boldsymbol{\Omega}_n \boldsymbol{W}_n \boldsymbol{S}_n)
(\boldsymbol{S}_n^\prime \boldsymbol{W}_n \boldsymbol{S}_n)^{-1},
$$
where we use the shorthand notation
$$
\textstyle
\boldsymbol{S}_n
=
\sum_g 
\boldsymbol{Z}_g^\prime   \boldsymbol{X}_g,
\qquad
\boldsymbol{W}_n
=
(
\sum_g 
\boldsymbol{Z}_g^\prime   \boldsymbol{Z}_g
)^{-1},
\qquad
\boldsymbol{\Omega}_n
=
\sum_g 
\boldsymbol{Z}_g^\prime \boldsymbol{\hat{\varepsilon}}_g  \boldsymbol{\hat{\varepsilon}}_g^\prime   \boldsymbol{Z}_g,
$$
and $\boldsymbol{\hat{\varepsilon}}_g = \boldsymbol{y}_g - \boldsymbol{X}_g\boldsymbol{{\vartheta}}_n$ are the residuals from the two-stage least-squares procedure.
We then have
$$
\boldsymbol{V}_n^{-1/2} (\boldsymbol{\vartheta}_n-\boldsymbol{\vartheta})
\overset{d}{\rightarrow} N(\boldsymbol{0},\boldsymbol{I}_4)
$$
as $n\rightarrow\infty$.

\section{Simulations} \setcounter{equation}{0}
The procedure was evaluated in a Monte Carlo experiment. We generated networks via the link formation process
\begin{equation*} 
(\boldsymbol{A})_{i,j} = 
\left\lbrace
\begin{array}{cl}
1 & \text{ if } \eta_i+\eta_j > c \\
0 & \text{ otherwise}
\end{array}	
\right. ,
\end{equation*}
where the $\eta_i$ are independent standard-normal variates and we set $c = -\sqrt{2}\Phi^{-1}(.25)$, for $\Phi$ the standard-normal distribution function. In this way, the unconditional link-formation probability is $.25$. We then drew $x_i\sim N(1,1)$ and generated outcomes from the full model, inducing endogeneity in link formation by generating
$$
\varepsilon_i = \varphi(\eta_i) + u_i, \qquad u_i \sim N(0,1),
$$
for different choices of the function $\varphi$. The parameters were set as $\alpha=0$, $\beta =1$, $\gamma=.5$ and $\delta =.5$. Data were generated for 250 groups, each consisting of 25 agents. Results are presented for the estimator of \cite{BramoulleDjebbariFortin2009} (TSLS-X) and for our proposal (TSLS-E). The former instruments $\boldsymbol{H}\boldsymbol{x}$ by itself and $\boldsymbol{H}\boldsymbol{y}$ by $\boldsymbol{H}^2\boldsymbol{x},\ldots, \boldsymbol{H}^4\boldsymbol{x}$. This approach is valid when $\varphi(\eta)$ does not depend on $\eta$. The latter instruments $\boldsymbol{H}\boldsymbol{x}$ by $\boldsymbol{Q}_1\boldsymbol{x}$  and $\boldsymbol{H}\boldsymbol{y}$ by $\boldsymbol{Q}_2\boldsymbol{x},\ldots, \boldsymbol{Q}_4\boldsymbol{x}$. We use two overidentifying moments to discipline the sampling distribution of the estimators---ensuring that their first two moments exist---so that we can meaningfully report on their bias and standard deviation (see, e.g., \citealt{Mariano1972}).

Table \ref{table:sim1} contains the bias and standard deviation of the estimators, the mean and standard deviation of the implied $t$-statistics, as well as the empirical rejection frequency of two-sided $t$-tests (at the $5\%$ significance level). We do not report results for the estimator of the intercept. Four different specifications for $\varphi$ were considered: (i) a constant, (ii) a linear function, (iii) an exponential function, and (iv) a sine function. All results were obtained over 5,000 Monte Carlo replications and all the variables were redrawn in each iteration.

\begin{table} \footnotesize
	\caption{Simulation results} \label{table:sim1}
\begin{tabular}{crcrccrcrcc}

\hline\hline

& \multicolumn{5}{c}{TSLS-X} & \multicolumn{5}{c}{TSLS-E} \\

 & \multicolumn{2}{c}{$\boldsymbol{\vartheta}_n-\boldsymbol{\vartheta}$} & 
\multicolumn{3}{c}{$\boldsymbol{V}_n^{-1/2}(\boldsymbol{\vartheta}_n-\boldsymbol{\vartheta})$}   & \multicolumn{2}{c}{$\boldsymbol{\vartheta}_n-\boldsymbol{\vartheta}$} & 
\multicolumn{3}{c}{$\boldsymbol{V}_n^{-1/2}(\boldsymbol{\vartheta}_n-\boldsymbol{\vartheta})$}  \\

\hline

 & \multicolumn{1}{c}{bias} & std & \multicolumn{1}{c}{mean} & std & rate   & \multicolumn{1}{c}{bias} & std & \multicolumn{1}{c}{mean} & std & rate \\

\multicolumn{11}{c}{$\varphi(\eta) = 0$} \\
								
$\beta$	 &
0.0000 & 0.0130 & 0.0005 &	1.0320 &	0.0578 &	-0.0002 &	0.0134 &	-0.0165 &	1.0286 &	0.0550 \\
$\gamma$ &	-0.0002 &	0.0414 &	-0.0077 &	1.0084 &	0.0542 &	-0.0002 &	0.1638 &	-0.2083 &	1.0287 &	0.0600 \\
$\delta$ &	0.0000 &	0.0203 &	0.0241 &	1.0047 &	0.0508 &	-0.0001 &	0.0610 &	0.2144 &	1.0235 &	0.0594 \\
\multicolumn{11}{c}{$\varphi(\eta) = \eta$} \\								
$\beta$ &	-0.0081 &	0.0181 &	-0.4580 &	1.0244 &	0.0816 &	0.0002 &	0.0180 &	0.0134 &	1.0049 &	0.0532 \\
$\gamma$ &	-0.0833 &	0.1336 &	-1.2585 &	2.0672 &	0.4930 &	-0.0003 &	0.1406 &	-0.1060 &	1.0261 &	0.0572 \\
$\delta$ &	-0.0833 &	0.0651 &	7.7740 &	3.1595 &	0.9596 &	-0.0004 &	0.0459 &	0.1075 &	1.0224 &	0.0548 \\
\multicolumn{11}{c}{$\varphi(\eta) = \exp(3\Phi(\eta))$} \\								
$\beta$ &	-0.0073 &	0.0661 &	-0.1140 &	1.0071 &	0.0538 &	0.0007 &	0.0669 &	0.0083 &	1.0009 &	0.0524 \\
$\gamma$ &	0.1873 &	0.4531 &	0.7775 &	1.8904 &	0.2462 &	-0.0079 &	0.5682 &	-0.0597 &	1.0186 &	0.0534 \\
$\delta$ &	0.1515 &	0.0611 &	8.1131 &	3.5276 &	0.9460 &	0.0013 &	0.1326 &	0.1113 &	1.0358 &	0.0626 \\
\multicolumn{11}{c}{$\varphi(\eta) = \sin(3\Phi(\eta))$} \\								
$\beta$ &	-0.0017 &	0.0130 &	-0.1336 &	0.9951 &	0.0496 &	-0.0003 &	0.0133	 & -0.0209 &	0.9899 &	0.0468 \\
$\gamma$ &	-0.0315 &	0.0436	 &-0.7777 &	1.0663 &	0.1406 &	-0.0139 &	0.1552 &	-0.2240 &	1.0319 &	0.0652 \\
$\delta$ &	0.0379 &	0.0200 &	2.4044 &	1.3197 &	0.6308 &	0.0055 &	0.0572 &	0.2364 &	1.0230 &	0.0624 \\

\hline\hline

\end{tabular}
\end{table}

The estimator of \cite{BramoulleDjebbariFortin2009} does well when link formation is exogenous. Otherwise, the coefficient estimates are biased, except for those of $\beta$. The latter observation can be explained by the fact that link formation is independent of the covariates in our design here. The (estimated) standard error (not reported) also tends to substantially underestimate the true variability in the point estimates. Together with the presence of bias, this implies that the $t$-statistics constructed from TSLS-X have a mean that is far from zero and a variance that greatly exceeds unity. Consequently, the $t$-test displays large overrejection rates.
Using instruments constructed from the leave-one-out networks delivers estimators that are virtually unbiased for all the designs in Table \ref{table:sim1}. The associated $t$-statistics have a mean that is close to zero and a standard deviation that is close to unity. Furthermore, the empirical rejection frequencies are close to their nominal size of $5\%$, and this for all parameters and for all designs. Hence, the normal approximation does well for TSLS-E.

The findings discussed here were confirmed in a larger set of Monte Carlo designs, where the $\eta_i$ were drawn from asymmetric distributions and network formation also depends on covariates. These results were similar in spirit to those reported here and, hence, are not discussed further here.






\footnotesize 

\setlength{\bibsep}{1pt} 
\bibliographystyle{chicago3}
\bibliography{bibliography}

\end{document}